\documentstyle[aps,floats,psfig,twocolumn]{revtex}
\def\gz{\ifmmode{Z\hskip -4.8pt Z}
    \else{\hbox{$Z\hskip -4.8pt Z$}}\fi} 

\newcommand{\be}{\begin{equation}}
\newcommand{\ee}{\end{equation}}
\newcommand{\bea}{\begin{eqnarray}}
\newcommand{\eea}{\end{eqnarray}}

\begin{document}
\tighten
\draft
\title{Nature of the insulating phases in the half-filled ionic
Hubbard model}

\author{A.~P. Kampf $^{a}$, M. Sekania $^{a}$, G.~I. Japaridze $^{a,b}$, and
Ph. Brune $^{a}$}
\address{
$^a$ Institut f\"ur Physik, Theoretische Physik III,
Elektronische Korrelationen und Magnetismus,\\
Universit\"at Augsburg, 86135 Augsburg, Germany\\
$^b$ Institute of Physics, Georgian Academy of Sciences,
Tamarashvili Str. 6, 380077, Tbilisi, Georgia}
\address{~
\parbox{14cm}{\rm
\medskip
We investigate the ground-state phase diagram of the one-dimensional ``ionic''
Hubbard model with an alternating periodic potential at half-filling by
numerical diagonalization of finite systems with the Lanczos and density
matrix renormalization group (DMRG) methods. We identify an
insulator-insulator phase transition from a band to a correlated insulator
with simultaneous charge and bond-charge order. The transition point is
characterized by the vanishing of the optical excitation gap while
simultaneously the charge and spin gaps remain finite and equal. Indications
for a possible second transition into a Mott-insulator phase are discussed.
}}

\maketitle

\section{Introduction}

For more than two decades the correlation induced metal-insulator transition
and its characteristics has been one of the challenging problems
in condensed matter physics \cite{Imada}. This metal-insulator transition is
often accompanied by a symmetry breaking and the development of long range
order \cite{Lee}. In one dimension this ordering can only be related to
the breaking of a discrete symmetry. Examples include commensurate charge
density waves (CDWs) and Peierls dimerization (bond-order wave, BOW) phenomena.
In contrast, the transition into the Mott insulating (MI) phase in one dimension
is not connected with the breaking of a discrete symmetry \cite{LiebWu}. In the
MI phase the {\it gapped charge} degrees of freedom are uniformly distributed in
the system, while the {\it gapless spin} degrees of freedom are described by
an effective $S=1/2$ Heisenberg chain \cite{Mattice}.

Due to the different symmetries of the CDW, BOW, and MI phases it is natural
to expect that these phases are mutually exclusive.
The extended Hubbard model at half-filling with an on-site ($U$) and a nearest
neighbor ($V$) Coulomb repulsion provides a prominent example with a
transition from a MI to a CDW insulator in the vicinity of the $U=2V$ line in
the phase diagram \cite{Emery}. Remarkably, the transition at weak coupling
may involve an intermediate BOW phase \cite{Nakamura,Campbell}. A similar
phase-diagram structure was recently also discovered for the Holstein Hubbard
model \cite{FehskeAHHM}. The tendency towards BOW order is even more profound
for the Hubbard model with an explicit bond-charge coupling where the CDW and
MI phases are often separated by a long range ordered Peierls dimerized phase
\cite{KampfJaparidze}.

In recent years particular attention has been given to another example for an
extension of the Hubbard model which includes a staggered potential term
\cite{Nagaosa86,Resta95,Fabrizio99,Gidopoulos99,Wilkens00,Pati00,Takada00,Qin00,Torio01}.
The corresponding Hamiltonian has been named the ``ionic Hubbard model'' (IHM);
in one dimension it is given by
\bea
H  = &-&t\sum_{i,\sigma}(1+(-1)^{i}\delta)\left(
c^{\dagger}_{i\sigma}c^{\phantom{\dagger}}_{i+1\sigma} + H.c.\right)\nonumber\\
&+& U \sum_i n^{\phantom{\dagger}}_{i\uparrow}n^{\phantom{\dagger}}_{i\downarrow}
+{\Delta\over 2}\sum_{i,\sigma} (-1)^i
            n^{\phantom{\dagger}}_{i\sigma}
 \,,
 \label{IonicHam}
\eea
where $c^{\dagger}_{i\sigma}$ creates an electron on site $i$ with spin
$\sigma$ and $n^{\phantom{\dagger}}_{i\sigma}=c^{\dagger}_{i\sigma}
c^{\phantom{\dagger}}_{i\sigma}$.
$\Delta$ is the potential energy difference between neighboring sites, and
$\delta$ a Peierls modulation of the hopping amplitude $t$. In the limit
$\Delta=\delta=0$, Eq. (\ref{IonicHam}) reduces to the ordinary
Hubbard model, the limit $\Delta=0$ and $\delta>0$ is called the
Peierls-Hubbard model, and the limit $\Delta>0$ and $\delta=0$ is usually
referred to as the IHM. In the following, we will focus mainly on the
effect of the on-site modulation $\Delta$, so we implicitly assume $\delta=0$
except where stated otherwise.

The IHM was first proposed and discussed almost 20 years ago in the context of
organic mixed-stack charge-transfer crystals with alternating donor ($D$) and
acceptor ($A$) molecules ($\dots D^{+\rho}A^{-\rho}\dots$)
\cite{Torrance81,Nagaosa86}. These stacks form quasi-1D insulating chains,
and at room temperature and ambient pressure are either mostly ionic
($\rho\approx 1$) or mostly neutral ($\rho\approx 0$)\cite{Torrance81}.
However, several systems undergo a reversible neutral to ionic phase
transition i.e. a discontinuous jump in the ionicity $\rho$ upon
changing temperature or pressure \cite{Mitani84}. Later the  IHM has been
used in a similar context to describe the ferroelectric transition in
perovskite materials such as BaTiO$_3$ \cite{Egami93,Ishihara94} or KNbO$_3$
\cite{Neumann92}.

The very presence of at least on transition in the ground state phase diagram
of the half-filled IHM model is easily traced by starting from the atomic
limit \cite{McConnell,Gidopoulos99}. For $t=0$, it is obvious that at
$U<\Delta$ the ground state of the IHM has two electrons on the odd sites, and
no electrons on the even sites corresponding to CDW
order with maximum amplitude. On the other hand, for $U>\Delta$ each site
is occupied by one electron with infinite spin degeneracy. Thus, for $t=0$ a
transition occurs at a
critical value $U_c=\Delta$. This transition is expected to persist for
finite hopping amplitudes $t>0$.

A renewal of interest in the IHM started with the bosonization analysis of
Fabrizio et al. (FGN) \cite{Fabrizio99}, where a {\it two-transition scenario}
for the ground-state phase diagram of the 1D IHM was
proposed. The key features of the FGN theory are the presence of an
{\it  Ising-type transition} from a CDW band-insulator phase at
$U<U^c_{ch}$ into a BOW phase at $U^c_{ch}<U<U^{c}_{sp}$, and a
continuous {\it Kosterlitz-Thouless like transition} into a MI phase at
$U >U^{c}_{sp}$. In this scenario the charge gap vanishes {\it only} at
$U=U^c_{ch}$ and the system might be ``metallic'' at this point. The second
transition at $U=U^{c}_{sp}$ is connected with the
closing of the spin gap, which is finite for all $U < U^{c}_{sp}$ and vanishes
for $U>U^{c}_{sp}$. Thus, the bosonization phase diagram
essentially supports the "exclusion principle" of the ground states.

Later on various attempts based on numerical tools have been performed to
verify the FGN phase diagram for the half-filled IHM. In particular, exact
diagonalization \cite{Gidopoulos99,Torio01}, valence bond
techniques \cite{Pati00}, quantum Monte Carlo \cite{Wilkens00}, and DMRG
\cite{Takada00,Qin00} was used. Unfortunately, conflicting results have so
far been reported in these studies regarding the nature of the transition
and the insulating phases, the possibility of two rather than one critical
point, or the appearance of BOW order.

Given the numerous unresolved issues we reinvestigate the ground-state
properties of the IHM using the exact diagonalization Lanczos
technique and the DMRG method.
We verify the presence of {\it at least one transition} at a critical
coupling $U_c(\Delta)$ from a band-insulator (BI) to
a correlated insulator (CI) phase. On finite systems the transition originates
from a ground-state level crossing with a change of the site-parity
eigenvalue, which implies the vanishing
of the optical excitation gap at $U_c$; Our DMRG results show that the spin
and charge gaps remain nevertheless {\it finite} and {\it equal} at the
transition. Above $U_c$ the charge and spin gaps split, the charge gap
increases, while the spin gap decreases and we identify
long-range BOW order with a spontaneous site-inversion symmetry breaking.
The existence of a second transition is not unambiguously resolved within the
accuracy of our DMRG data. Yet, the scaling of the BOW order parameter changes
qualitatively with increasing $U$ indicative for a possible second smooth
transition point where the spin gap closes.

We show that at $U<U_{c}(\Delta)$ the CDW-band insulator phase is realized.
In this phase BOW and spin density wave (SDW) correlations are strongly
suppressed, and the {\em spin and charge gaps are equal} and finite. The
characteristic feature of the ground-state phases for $U>U_c(\Delta)$ is
the coexistence of long range CDW order with either a long range BOW or
algebraically decaying BOW and SDW correlations.

\section{Symmetry analysis}

A good starting point for understanding the existence of a phase transition in
the IHM is to study the symmetry of the model manifestly seen in the
limiting cases $U \ll \Delta, t$ and $U \gg \Delta, t$.
The IHM is invariant with respect to {\it inversion} at a site and
{\it translation} by two lattice sites. If we denote the site inversion
operator by $P$, defined through
\begin{equation}
 Pc^{\dagger}_{i\sigma}P^{\dagger}=c^{\dagger}_{L-i\sigma}
 \quad\mathrm{for} \;i=0, \cdots, L-1
 \,\,\,,
\end{equation}
and ${\hat T}_j$ for a translation
by $j$ sites, then any nondegenerate eigenstate $|\psi_n\rangle$ of $H$ must
obey $P|\psi_n\rangle=\pm|\psi_n\rangle$ and
${\hat T}_2|\psi_n\rangle=|\psi_n\rangle$. Because $[H,{\hat T}_1]\neq 0$, any
non-degenerate eigenstate $|\psi_n\rangle$ of $H$ is not an eigenstate
of ${\hat T}_1$.

For the half-filled Hubbard model ($\Delta=\delta=0$) the ground state
has $P=+1$ only for $U=0$, and $P=-1$ for any $U>0$ \cite{Gidopoulos99}.
However, in the IHM the phase transition from a renormalized BI to a CI occurs
at some finite $U_c>0$. This suggests
that the parity of the ground state remains even not only for $U=0$, but for
all $U<U_c$. At $U_c$, a ground-state level crossing occurs on finite chains,
as confirmed by exact diagonalization studies (see below), connected with a
site-parity change.

For $U=0$ the ground state at half-filling is a CDW-BI. The alternating
potential defines two sublattices, doubling the unit cell and opening up a
band gap $\Delta$ for $U=0$ at $k=\pm\pi/2$. The elementary spectrum consists
of particle-hole excitations over the band gap. The charge ($\Delta_C $) and
spin ($\Delta_S $)  excitation gaps are equal: $\Delta_C=\Delta_S=\Delta $.
We consider a system to be a BI when the criterion $ \Delta_S=\Delta_C$
holds, where the spin and the charge gaps are given by
\begin{eqnarray}
 \Delta_S &=& E_0(N=L,S_z=1) \nonumber\\
          &-& E_0(N=L,S_z=0) \, , \nonumber\\
 \Delta_C &=& E_0(N=L+1,S_z=1/2) \nonumber\\
          &+& E_0(N=L-1,S_z=1/2) \nonumber\\
          &-& 2E_0(N=L,S_z=0),
\label{Gaps1}
\end{eqnarray}
respectively. $E_0(N,S_z)$ is the ground-state energy, $L$ the system length,
$N$ the number of electrons, and $S_z$ the $z$-component of the total spin.
As we show below the BI phase is realized in the ground state of the IHM at
$U<U_c$.

In the strong-coupling limit $U \gg \Delta, t$, the low-energy physics of the
IHM is described by the following effective Heisenberg spin model
\cite{Nagaosa86}
\begin{equation}
 H_{eff}=J\sum_{i}{\bf S}_i\cdot{\bf S}_{i+1}
         +J'\sum_{i}{\bf S}_i\cdot{\bf S}_{i+2}\, .
\label{SpinHam}
\end{equation}
In Eq. (\ref{SpinHam}) the exchange couplings are given by
\begin{eqnarray}
 \nonumber
 J      &=& \frac{4t^2}{U}\left[\frac{1}{1-x^2}
            -\frac{4t^2}{U^2}\frac{1+4x^2-x^4)}{(1-x^2)^3}\right]\, ,\\
 J'     &=& \frac{4t^4}{U^3}\frac{(1+4x^2-x^4)}{(1-x^2)^3}
 \label{JCouplings}
 \,,
\end{eqnarray}
where $x=\Delta/U$. This result (\ref{SpinHam}) implies that in the
strong-coupling limit of the
IHM the low-energy physics is qualitatively similar to that of the Hubbard
model, with modified exchange coupling constants $J$ and $J'$. For
next-nearest neighbor couplings $J'<0.24J$ the spin gap vanishes
\cite{Haldane}. The coupling constants
(\ref{JCouplings}) satisfy the condition $J'<0.24J$ at least for $U>3.6t$ for
$\Delta\le t$ and $U>3.6\Delta$ for $\Delta>t$.

The effective spin model (\ref{SpinHam}) is invariant with respect
to translations by {\it one} lattice spacing, whereas the original IHM is
invariant only with respect to translations by {\it two} lattice spacings.
However, the doubling of the unit cell is ensured due to the charge
degrees of freedom. For the standard Hubbard model at arbitrary
$U\neq 0$ the number of doubly occupied sites $D$ in the ground state is
finite. The exact Bethe-ansatz solution tells that $D$ scales as $(t/U)^{2}$
in the strong coupling limit and is given
by \cite{CarmeloBaeriswyl}
\be\label{Dba}
D = \sum_i \langle n^{\phantom{\dagger}}_{i\uparrow}
n^{\phantom{\dagger}}_{i\downarrow}\rangle
\simeq N A (t/U)^{2}[1+ {\cal O}\left((t/U)^{2}\right)],\nonumber
\ee
where $A=4\ln 2$. Contrary to the  Hubbard model, where
doublons are equally distributed on all sites of the system, the non-equivalence
of sites in the IHM leads to different probabilities for finding
a doublon on even or odd sites. Since doublons are spin singlets, their
distribution is not influenced by the spin fluctuations.
Since the energy of doublons on even and odd sites differ in $\Delta$
and assuming  the scaling for the density of doublons as in Eq. (\ref{Dba}),
one easily obtains for the amplitude of the ionicity induced CDW in the strong
coupling limit
\bea\label{IndCDW}
\frac{1}{N}\left(D_{\it odd} - D_{\it even}\right) \simeq
A_{1}\frac{t^{2}}{U^{2}}\left[\frac{1}{(1-x)^{2}}
- \frac{1}{(1+x)^{2}}\right]\nonumber \\
= 4A_{1}\frac{t^{2}}{U^{2}}\frac{x}{(1-x^{2})^{2}}[1+
{\cal O}\left((t/U)^{2}\right)]
\eea
where $A_{1}$ is a constant of order unity. Thus, although the
effective spin Hamiltonian has a higher symmetry than
the original model from which it was derived, the translational symmetry
of the IHM is recovered due to the long range CDW pattern arising from the
staggered doublon and holon distribution.

\section{Exact diagonalization results}

In order to explore the nature of the spectrum and the phase transition,
we have diagonalized numerically small systems by the Lanczos method
\cite{Lanczos50} similarly to earlier exact diagonalization calculations
\cite{Resta95,Gidopoulos99}. The energies of the few lowest eigenstates were
obtained for finite chains with $L=4n$ and periodic boundary conditions or
$L=4n+2$ with antiperiodic boundary conditions.

We first analyze short chains; for chain lengths $L\le 16$ finite-size effects
do not change the qualitative behavior discussed below. In Fig. 1,
the lowest eigenenergies of the IHM for $\Delta=0.5t$,
$L=8$ and periodic boundary condition are shown as a function of $U$.
At $U=1.3t$, a level crossing of the two lowest eigenstates occurs. A
non-degenerate eigenstate of the IHM has a well defined site-parity, so a
ground-state level-crossing transition necessarily corresponds to a change of
the site-parity eigenvalue.

For $U=0$, the IHM is easily diagonalized in momentum space by introducing
fermionic creation operators $\gamma^{\dagger}_{k\sigma b}$
with a band index $b=1,2$ for the lower and upper bands, respectively, with the
dispersion $E_{1/2}(k)=\pm\sqrt{4\cos^2(k)+(\Delta/4)^2}$ for momenta
$-\pi/2<k\le\pi/2$. For $U=0$ the first two degenerate
excited states at half-filling always have negative site parity, because the
ground state has $P=+1$, and the
operator $\gamma^{\dagger}_{q\sigma 2}\gamma^{\phantom{\dagger}}
_{q\sigma 1}$ with $q=\pi/2$ obeys
\begin{equation}
P\gamma^{\dagger}_{q\sigma 2}
 \gamma^{\phantom{\dagger}}_{q\sigma 1}
=-\gamma^{\dagger}_{q\sigma 2}
  \gamma^{\phantom{\dagger}}_{q\sigma 1}P
 \,.
\end{equation}
The first two excited states shown in Fig. 1 are the spin
singlet ($S=0$, $S_z=0$) and triplet excitations ($S=1$, $S_z=0$), created
from the ground state by applying the operators
\begin{eqnarray}
 \nonumber
 &&\frac{1}{\sqrt{2}}\left(
 \gamma^{\dagger}_{q\uparrow 2}
 \gamma^{\phantom{\dagger}}_{q\uparrow 1}
 -\gamma^{\dagger}_{q\downarrow 2}
  \gamma^{\phantom{\dagger}}_{q\downarrow 1}\right)\, ,\\
 &&\frac{1}{\sqrt{2}}\left(
 \gamma^{\dagger}_{q\uparrow 2}
 \gamma^{\phantom{\dagger}}_{q\uparrow 1}
 +\gamma^{\dagger}_{q\downarrow 2}
  \gamma^{\phantom{\dagger}}_{q\downarrow 1}\right)
 \,,
\end{eqnarray}
respectively. Thus both excited states have total momentum $k_{tot}=0$ and
negative site parity. For $U>0$, these degenerate excited states split in
energy. Exact diagonalization of finite IHM rings therefore
identifies one critical $U_c>0$, separating a BI with $P=+1$ at $U<U_c$ from
a CI with $P=-1$ for $U>U_c$.

\begin{figure}
\centerline{\psfig{file=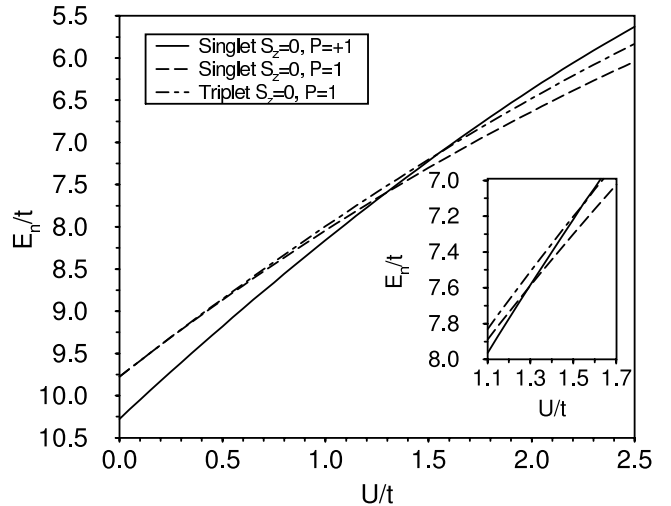,width=85mm,silent=}}
\vspace{4mm}
\caption[LE]{Lowest energy eigenvalues of the IHM at
half-filling for $L=8$ sites, periodic boundary conditions and $\Delta=0.5t$.}
\label{fig:Energies}
\end{figure}

\section{DMRG results}
\subsection{Excitation gaps}

In order to access the transition scenario in the long chain-length limit, we
have studied chains up to L=512 using the DMRG method
\cite{White92,White93,Peschel99}. The fact that the transition at $U_c$ is
connected to a change in inversion symmetry
requires some caution when open boundary conditions (OBC) are used in DMRG
studies. For OBC and $L=2n$ the IHM is not reflection symmetric at any site.
Thus, the ground state does not have a well defined site parity, and the
level-crossing transition is absent.
To overcome this problem, one might try to use chains with OBC and
an {\it odd} number of sites $L=2n+1$, since the Hamiltonian in this case is
reflection symmetric with respect to the site $i_c$ in the center of the chain,
and a site inversion operator is well defined by
\begin{equation}
Pc^{\dagger}_{i_c\sigma}P^\dagger
=c^{\dagger}_{L-1-i_c\sigma}\quad i=0,...,L-1\,\, .
\end{equation}
To test whether this is an improved choice we have calculated the site parity
of the ground state for $U=0$ analytically for different chain lengths
$L=2n+1$ and found
\begin{equation}
 P|\psi_0\rangle=(-1)^n|\psi_0\rangle
 \,.
\end{equation}
On the other hand, if one extends the idea of Gidopoulos {\it et al.}
\cite{Gidopoulos99} for the determination of the site parity to chains with
$L=2n+1$ for $U\gg t$, one obtains
\begin{equation}
 P|\psi_0\rangle = (-1)^{\left[\sum^{L-1}_{m=1}m\right]}|\psi_0\rangle
                 = (-1)^{n}|\psi_0\rangle
 \,.
\end{equation}

Thus, the parity eigenvalue of the ground state is the same at $U=0$ and
$U\gg t$ for a given chain length, and no level crossing occurs. Due to
the fact that the sharp transition at a well defined $U_c$ does not exist in
the finite-chain results for OBC, the extrapolation is a rather subtle
problem, since a sharp transition feature has to be identified from the
extrapolation of smooth curves. This requires the use of quite long chains in
the critical region.

\begin{figure}[tbh]
\centerline{\psfig{file=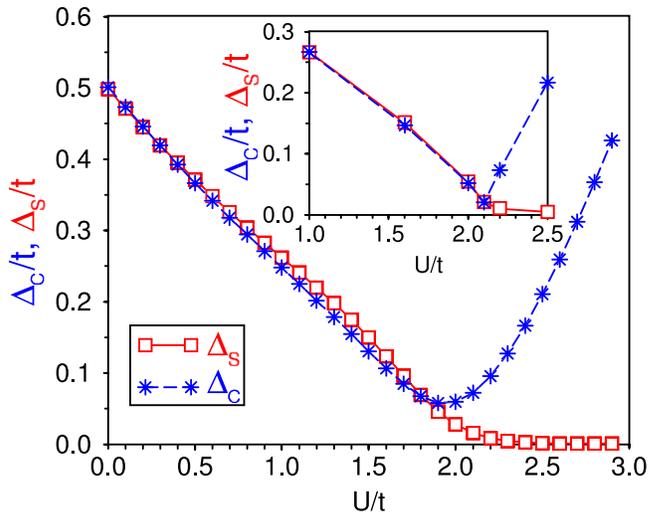,width=85mm,silent=}}
\vspace{4mm}
\caption{Results for the spin ($\Delta_S$) and charge ($\Delta_C$) gaps of the
IHM at half-filling with $\Delta=0.5t$ as a function of $U$.
Energies were obtained by DMRG calculations on open chains with
$L=\{30,40,50,60\}$ (main plot) and up to $L=512$ (inset), and
extrapolated to the limit of infinite chain length.}
\label{fig:UMuGaps}
\end{figure}

In Fig. \ref{fig:UMuGaps} extrapolated results are shown for the spin and
charge gaps $\Delta_S$ and $\Delta_C$, respectively.
Calculations were performed with OBC for chains of lengths
$L=\{30,40,50,60\}$, and additionally up to $L=512$ in the transition region
around the estimated $U_c$. We assume a scaling behavior of $\Delta_C$ and
$\Delta_S$ of the form \cite{Noack00}
\begin{equation}
\Delta_i(L)=\Delta^{\infty}_i+\frac{A_i}{L}+\frac{B_i}{L^2}\, ,
\label{GapScaling}
\end{equation}
where $i\in\{S,C\}$. The extrapolation for $L\rightarrow\infty$ is then
performed by fitting this polynomial in
$1/L$ to the calculated finite-chain results. We note that different
finite-size scaling formulas were proposed in the literature mainly when
periodic or antiperiodic boundary conditions were used \cite{Gidopoulos99}.

As can be seen from the main plot in Fig. \ref{fig:UMuGaps}, extrapolating the
results for $L=\{30,40,50,60\}$ does indeed not give a sharp transition
behavior. As illustrated in the inset, adding results for $L$ up
to $512$ in the critical region changes the picture considerably. Within
numerical accuracy the charge and spin gaps remain equal and finite up to a
critical $U_c\approx 2.1t$, where a sharp kink for $\Delta_C$ is observed.
Importantly, $\Delta_C$ does not close at the critical point. We emphasize that
the magnitude of $\Delta_C$ and $\Delta_S$ at the transition point is
sufficiently larger than our numerical uncertainty in the finite size scaling
analysis and therefore allows for a safe conclusion. $\Delta_C=\Delta_S>0$ at
the transition is in fact not in conflict with an underlying ground-state level
crossing. If the ground states of the different site-parity sectors become
degenerate, the only rigorous consequence is the closing of the
optical excitation gap. The selection rules for optical excitations allow only
for transitions between states of different site-parity. Furthermore, optical
transitions occur within the same particle number sector. The optical gap is
therefore by definition distinct from the charge gap Eq. (2) which involves
the removal or the addition of a particle. The critical point $U_c$
of the IHM has the remarkable peculiarity that the optical gap closes while
$\Delta_C$ remains finite. Above $U_c$ the charge and spin gaps split
indicating that the corresponding insulating phase is no longer a BI.
$\Delta_S$ continuously decreases with increasing $U$ and becomes unresolvably
small above $U\sim 2.5t$ within the achievable numerical accuracy.

The result, that $\Delta_C$ and $\Delta_S$ remain finite and equal at the
transition is in agreement with the data obtained by Qin {\it et al.}
\cite{Qin00}. These authors performed DMRG calculations for the IHM with
$\Delta=0.6t$ for open chains up to $L=600$ sites. They observed a surprising
non-monotonic scaling behavior of $\Delta_S$ with $L$ for values
of $U$ close to the critical $U_c$, i.e. for chain lengths $L>300$ $\Delta_S$
started to increase again. It remains unclear whether this is due to loss of
DMRG accuracy with increasing chain lengths and keeping a fixed number of
states in the DMRG algorithm. In contrast, our data always show a monotonous
scaling with $1/L$. DMRG calculations for the IHM with $\Delta=t$ have also been
performed by Takada and Kido for chains up to $L=400$ sites \cite{Takada00}. The
authors interpret their results in the region close to $U_c$ in favour of a
two-transition scenario similar to that of FGN \cite{Fabrizio99}. However, as
we will show below the accuracy of the currently available DMRG data is not
enough to provide a stringent argument in favour of this interpretation.

\begin{figure}[thb]
\centerline{\psfig{file=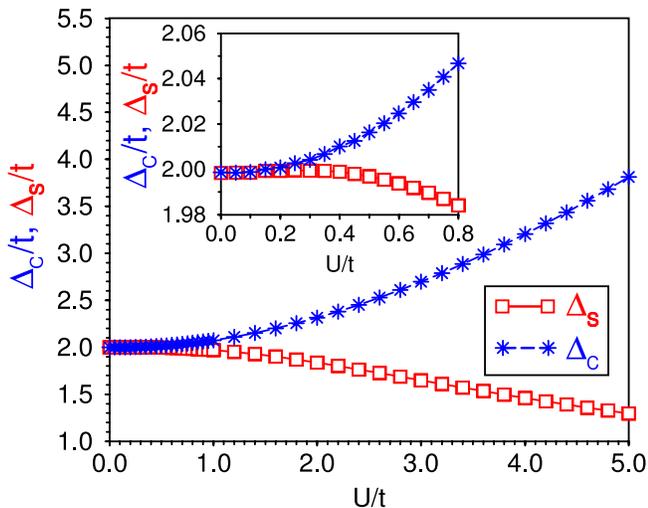,width=85mm,silent=}}
\vspace{4mm}
\caption{Results for the spin ($\Delta_S$) and charge ($\Delta_C$) gaps vs.
$U$ of the
Peierls-Hubbard model at half-filling with a modulation of the hopping
amplitude $\delta=0.5$. Energies were obtained by DMRG
calculations on open chains with
$L=\{30,40,50,60\}$.}
\label{fig:GapsPeirHub}
\end{figure}

For comparison we show in Fig. \ref{fig:GapsPeirHub} the spin and charge gaps
versus $U$ in the Peierls Hubbard model. As we observe this model is distinctly
different from the IHM. This is also a band-insulator at $U=0$, but in contrast
to the IHM has $\Delta_C>\Delta_S>0$ for any value $U>0$, i.e. the {\it phase
transition from the Peierls band-insulator to the correlated insulator occurs at
$U_c=0$}. So although the Peierls and the ionic BI for $U=0$ similarly possess
an excitation gap at the Brillouin zone boundary, applying a Coulomb $U$ leads
to distinctly different behavior in both cases. The origin of the different
behaviors must be traced to the fact that the Hubbard interaction and the
ionic potential compete locally on each site, while the Peierls modulation of
the hopping amplitude tends to move charge to the bonds between sites,
thereby avoiding conflict with the Hubbard term.

The spin-Peierls physics of the Peierls Hubbard model at large $U$ evolves
smoothly with decreasing $U$ into the physics of a spin-gapped CI-BOW state
in the weak-coupling limit, which is characterized by long-range staggered
bond-density correlations
\begin{eqnarray}
g_b(r)&=&{1\over L}\sum_{i}\,\langle\psi_0|\,b(i)b(i+r)\,|\psi_0\rangle
\,, \\
b(i)&=&\sum_\sigma\left( c^{\dagger}_{i\sigma}c^{\phantom{\dagger}}_{i+1\sigma}
                     +H.c.\right)\,.
\label{BOWCor}
\end{eqnarray}
For $U\gg t$ the low-energy physics of the Peierls Hubbard model
is described by the spin-Peierls Heisenberg Hamiltonian with a staggered
exchange interaction and a dimerization induced spin gap \cite{Bulaevskii}.

\subsection{Correlation functions}

The important question remains about the nature of the insulating phase of the
IHM for $U>U_c$. To further analyze the BI and CI phases below
and above $U_c$, we have evaluated site- and bond-charge distribution
functions as well as spin-spin correlation functions.
\begin{figure}[tbh]
\centerline{\psfig{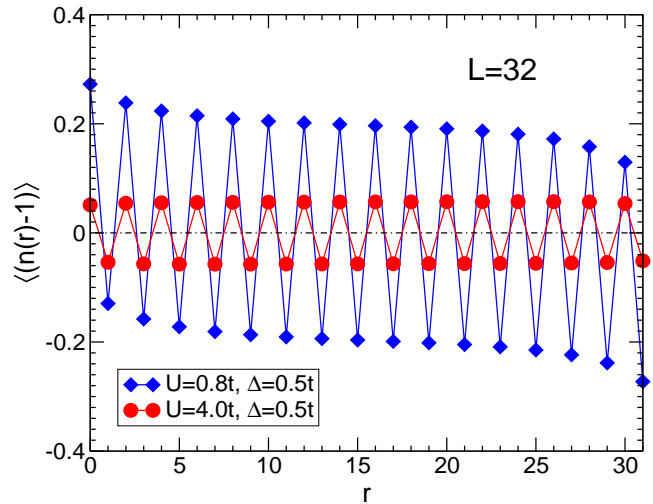}}
\vspace{4mm}
\caption{Electron density distribution in the ground state of the IHM for
$\Delta=0.5t$ and $U=0.8t$ (diamonds) and $U=4t$ (circles). Results were
obtained by  DMRG calculations on an open $L=32$ chain. }
\label{fig:NewCDW}
\end{figure}
\begin{figure}[tbh]
\centerline{\psfig{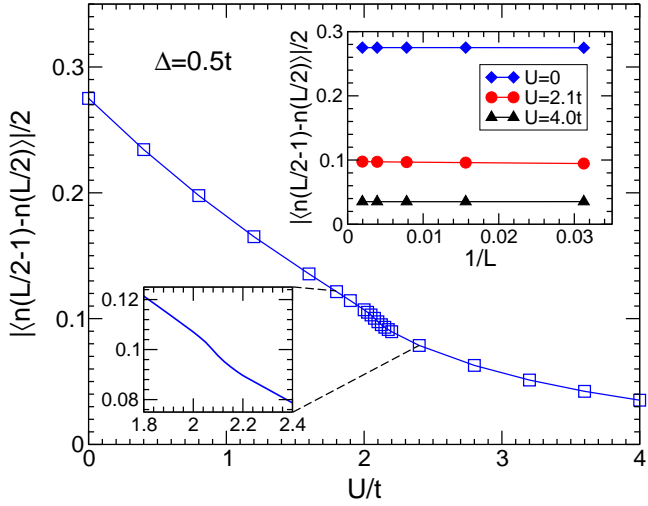}}
\vspace{4mm}
\caption{Staggered charge density component vs. $U$, for $\Delta=0.5t$ and
$L=512$ (main plot). Its scaling behavior, for $U=0$ (diamonds),
$U=2.1t$ (circles) and $U=4.0t$ (triangles) is shown in the inset.}
\label{fig:NewCDW_AMP}
\end{figure}
In Fig. \ref{fig:NewCDW} we show the charge distribution $\langle 0|(
n(r)-1)|0\rangle$ ($n(r)=n_{r,\uparrow}+n_{r,\downarrow}$)  in the ground state
$|0\rangle$ for the IHM at $U=0.8t<U_c$ and $U=4t>U_c$ for a $L=32$ chain.
The alternating pattern in the density distribution is well pronounced not
only in the BI but also in the CI phase far beyond the critical point at
$U\gg U_{c}$. For the $L=32$ chain the CDW is well established at distances
$l\sim L/2$ even at $U=4t$ and its amplitude remains almost unchanged in the
finite-size scaling analysis (see upper inset in Fig. \ref{fig:NewCDW_AMP}).
The main plot in Fig. \ref{fig:NewCDW_AMP} shows that the staggered
component of the charge density decreases smoothly with increasing $U$.
Close above the transition point near $U=2t$ one observes an anomaly,
i.e. a slight decrease of the CDW amplitude accompanied by a change in
curvature; this anomaly can be clearly identified in the enlargement shown
in the lower inset. The anomaly will find a natural explanation in the
discussion below. Our numerical data show that the alternating pattern in the
electron density distribution in the IHM remains for arbitrary finite $U$. Thus,
the ionic potential enforces long range CDW order for all interaction strengths.
\begin{figure}
\centerline{\psfig{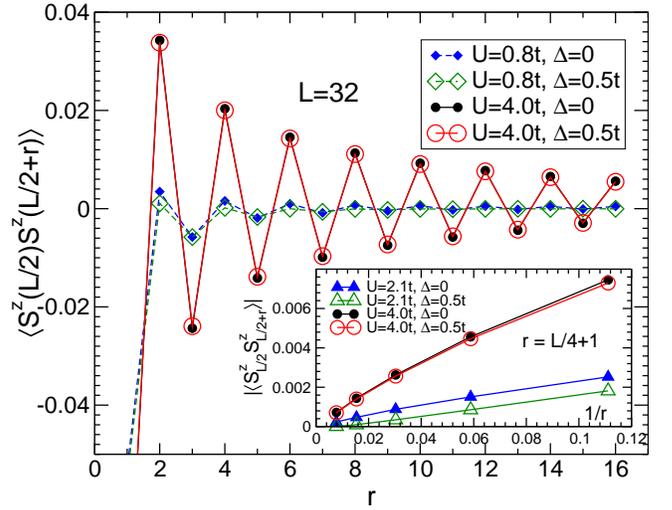}}
\vspace{4mm}
\caption{Spin-spin correlation function in the ground state of the IHM for
$\Delta=0.5t$ (open symbols) and the Hubbard model ($\Delta=0$) (full symbols)
at $U=0.8t$ (diamonds) and $U=4t$ (circles). Chain length $L=32$ (main plot).
Inset: Scaling of the spin-spin correlation function at $r=L/4+1$ for $U=2.1t$
(up and down triangles) and $U=4t$ (circles) for the IHM (open symbols) and the
Hubbard model (full symbols)}
\label{fig:NewSDW}
\end{figure}

Fig. \ref{fig:NewSDW} shows the DMRG results for the spin-spin correlation
function $\langle 0|S^{z}(L/2)S^{z}(L/2+r)|0\rangle$ for the IHM at
$U=0.8t<U_c$ and $U=4t>U_c$ in comparison with the Hubbard chain. In the BI
phase at $U=0.8t$ the SDW correlations are quickly suppressed after a few
lattice spacings. At $U=4t$ the amplitude of the SDW correlations in the CI
phase of the IHM is slightly reduced in comparison to the Hubbard model at the
same value of $U$. However, the large distance behaviors of the spin
correlations in the CI phase of the IHM and the MI phase of the Hubbard model
are quite similar and become almost indistinguishable (see the data in the
inset for $U=4t$). On the other hand, the long distance behavior of the
spin-spin correlation function at $U=2.1t$ and  $\Delta=0.5t$, i.e. close above
the transition, manifestly supports the finiteness of the spin gap (see inset).
This may be viewed as an indication for existance of two different phases above
$U_c$. But we cautiously point out that it is hard to judge on the
persistence or vanishing of $\Delta_S$ far above $U_c$ in the CI phase from
the finite chain spin correlators alone.

\begin{figure}[tbh]
\centerline{\psfig{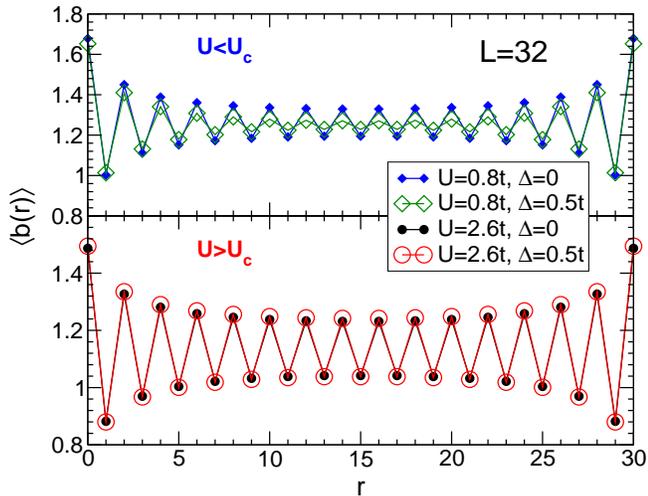}}
\vspace{4mm}
\caption{Bond-charge density of the IHM for
$\Delta=0.5t$ (open symbols) and the Hubbard model (full symbols) at $U=0.8t$
(diamonds) and $U=2.6t$ (circles).}
\label{fig:NewBDW}
\end{figure}

To address the BOW ordering tendencies in the CI phase we have calculated the
ground-state distribution of the bond-charge density Eq. (\ref{BOWCor}).
Fig. \ref{fig:NewBDW} shows the results of the DMRG calculations for the $L=32$
IHM and the Hubbard chain at $U=0.8t$ and $U=2.6t$. The boundary effect
for an open chain is strong and leads to a modulation of the bond density
already for the pure Hubbard model. Interestingly, the same behavior was also
observed previously for the bond expectation value $\langle{\bf S}_i\cdot
{\bf S}_{i+1}\rangle$ in open antiferromagnetic Heisenberg spin-$1/2$ chains
\cite{White93}. The detailed comparison with the Hubbard chain in
Fig. \ref{fig:NewBDW} reveals, that the ionic potential leads to a reduction
of the bond-density oscillations at $U<U_{c}$, while in the CI phase at
$U>U_{c}$ their amplitude slightly increases. The enhancement of BOW
correlations above $U_c$ must simultaneously weaken the CDW amplitude. This
naturally explains the slight downward curvature near $U_c$ in the staggered
charge density component shown in Fig. \ref{fig:NewCDW_AMP}.

\begin{figure}[tbh]
\centerline{\psfig{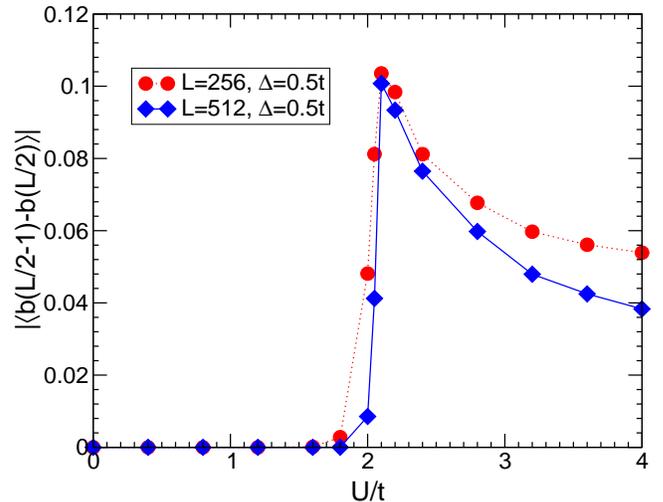}}
\vspace{4mm}
\caption{Staggered bond-density component vs. $U$ near the center of the
$L=256$ (circle, dotted line) and $L=512$ (diamonds, solid line) chain.}
\label{fig:BDW_U}
\end{figure}

In order to explore the possibility towards true long range BOW ordering in
the IHM above $U_c$ we plot in Fig. \ref{fig:BDW_U} the staggered bond-density
versus $U$ in the center of long, open chains with $L=256$ and $L=512$. In the
BI phase this quantity is essentially zero. At the transition point the
staggered component of the bond density increases rapidly, and on further
increasing $U$ it starts to decrease smoothly. However, the staggered bond
density remains finite for any $U>U_c$ on these long but finite chains and
vanishes only in the limit $U\rightarrow \infty$.

\begin{figure}[tbh]
\centerline{\psfig{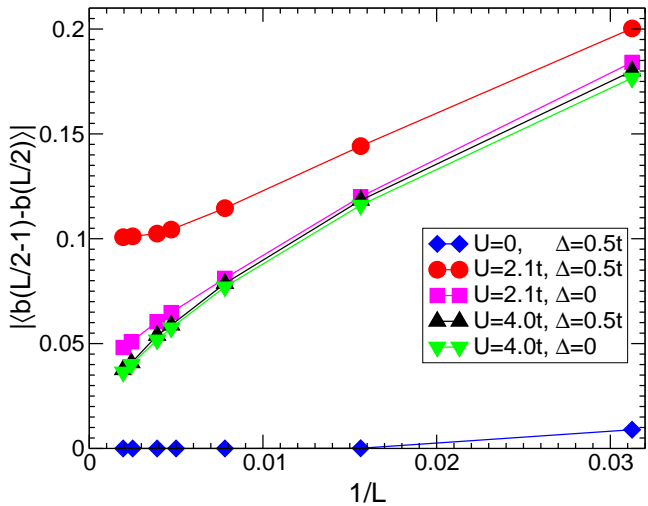}}
\vspace{4mm}
\caption{Scaling behavior of the staggered bond-density component in the IHM
near the center of the chain for $U=0$ (BI, diamonds), $U=2.1t$ (CI close above
the transition, circles), $U=4.0t$ (CI far above the transition,
up triangles) and in the Hubbard model for $U=2.1t$ (squares) and $U=4.0t$
(down triangles).}
\label{fig:BDW_SCA}
\end{figure}

Naturally it is necessary to perform a finite size scaling analysis for this
quantity. Fig. \ref{fig:BDW_SCA} shows
its chain length scaling behavior for $U=0$, $U=2.1t$, and $U=4t$. Two
conclusions can be drawn from these results: In the absence of the
interaction the staggered bond density clearly extrapolates to zero - as
expected for a conventional BI. In the CI phase close above the
transition the upward curvature of the staggered bond density vs. $1/L$ points
to a finite value in the infinite chain length limit, i.e. long range BOW
order. However, the scaling behavior changes in the CI phase far above the
transition point. Interestingly, the scaling behavior in the IHM far above
$U_c$ starts to reseamble the results for the pure Hubbard model, for which
the staggered bond density has to vanish in the thermodynamic limit.
The qualitative change in the scaling behavior of the staggered bond density
may again be viewed indicative for a possible second phase transition.
Summarizing the results for the SDW and BOW correlations we conclude that there
is evidence for two phases for $U>U_c$. Close above $U_c$ long range BOW order
develops in the ground state of IHM while far above the transition point
the coorrelation functions become almost identical to those of the Hubbard
model. Yet, long range CDW order exists for all $U$. A precise location of
a second transition point is however not possible from the currently awailable
DMRG data.

\section{Conclusions}

From the finite chain DMRG studies and finite  size scaling analyses we draw
the following conclusions for the ground-state phase diagram of the IHM: the
ionic potential leads to long range CDW order for all interaction strengths.
The data resolve one transition point from the BI to the CI phase. Remarkably,
at the transition $\Delta_C=\Delta_S$ and both remain finite. Close above the
transition, i.e. for $0<(U/U_c)-1\ll 1$, we identify a clear signal for
long range BOW order. With increasing $U$ above $U_c$ the finite size scaling
behavior of the staggered bond density and spin-spin correlation function changes
qualitatively and approaches the scaling behavior of the Hubbard model. With the
current chain length and DMRG accuracy limitations it is not possible to precisely
identify and locate the second transition point. The phase with BOW order
necessarily has a finite spin excitation gap. If BOW seizes to exist above a
second critical value of U, the spin gap has to vanish simultaneously leading to
identical large distance decays of SDW and BOW correlations.

The insulator-insulator transition at $U_c(\Delta)$ on finite periodic
chains results from a ground-state level crossing of the two site-parity
sectors. The optical excitation gap therefore has to vanish at $U_c$;
remarkably, the DMRG data reveal that at the critical point $\Delta_C$ ans
$\Delta_S$ remain both finite and equal. This itself clearly indicates the
existanse of a CI phase with $\Delta_C>\Delta_S>0$ which originates from
the appearance of long range staggered bond density order.
The distinction between the optical and the charge gap is therefore of key
importance for the structure of the insulating phases and the phase transitions
of the IHM. The investigation of the optical conductivity in the critical
region is therefore a demanding task for future work on the complex physics
of the ionic Hubbard model.

\acknowledgements

We thank R. Noack, B. Normand, H. Fehske, and T. Vekua for helpful
discussions. This work was supported by the Deutsche Forschungsgemeinschaft
(DFG) through SP 1073. APK also acknowledges support through
Sonderforschungsbereich 484 of the DFG. GIJ and MS acknowledge support by the
SCOPES grant N 7GEPJ62379.

\end{document}